# The space cold atom interferometer for testing the equivalence principle in the China Space Station

Meng He[1,2], Xi Chen[1,*], Jie Fang[1], Qunfeng Chen[1], Huanyao Sun[1], Yibo Wang[1], Jiaqi Zhong[1,3], Lin Zhou[1,3], Chuan He[1], Jinting Li[1,2], Danfang Zhang[1,2], Guiguo Ge[1,2], Wenzhang Wang[1,2], Yang Zhou[1,2], Xiao Li[1], Xiaowei Zhang[1], Lei Qin[1], Zhiyong Chen[1], Rundong Xu[1], Yan Wang[1], Zongyuan Xiong[1], Junjie Jiang[1,2], Zhendi Cai[1,2], Kuo Li[5], Guo Zheng[5], Weihua Peng[5], Jin Wang[1,3,4,†] and Mingsheng Zhan[1,3,4,‡]

[1] *State Key Laboratory of Magnetic Resonance and Atomic and Molecular Physics, Innovation Academy for Precision Measurement Science and Technology, Chinese Academy of Sciences, Wuhan 430071, China*

[2] *University of Chinese Academy of Sciences, Beijing 100049, China*

[3] *Hefei National Laboratory, Hefei, 230094, China*

[4] *Wuhan Institute of Quantum Technology, Wuhan 430206, China*

[5] *Wuhan Zmvision Technology Co., Ltd., Wuhan 430070, China*

Corresponding authors: [*]chenxi@apm.ac.cn

[†]wangjin@apm.ac.cn

[‡]mszhan@apm.ac.cn

**Abstract**: The precision of the weak equivalence principle (WEP) test using atom interferometers (AIs) is expected to be extremely high in microgravity environment. The microgravity scientific laboratory cabinet (MSLC) in the China Space Station (CSS) can provide a higher-level microgravity than the CSS itself, which provides a good experimental environment for scientific experiments that require high microgravity. We designed and realized a payload of a dual-species cold rubidium atom interferometer. The payload is highly integrated and has a size of 460 mm × 330 mm × 260 mm. It will be installed in the MSLC to carry out high-precision WEP test experiment. In this article, we introduce the constraints and guidelines of the payload design, the compositions and functions of the scientific payload, the expected test precision in space, and some results of the ground test experiments.

## 1 Introduction

General relativity (GR) is a cornerstone of modern physics. Einstein's equivalence principle (EEP) is one of the assumptions of GR and consists of three parts: the weak equivalence principle (WEP), local Lorentz invariance (LLI), and local position invariance (LPI). The WEP states that two objects with different materials and structures have the same acceleration in the same gravitational field.

GR describes gravity as a spacetime geometry, but gravity cannot be quantized by quantum mechanics (QM). Many new theories (loop quantum gravity, noncommutative geometry, and the fifth force) [1-3] have been proposed to solve the incompatibility between GR and QM. These theories generally require violation of the WEP. Therefore, high-precision tests of the WEP are important for identifying these new theories and searching for new physics.

The WEP test is usually carried out by measuring the accelerations of two different objects in the same gravitational field. The inequality of the measured accelerations indicates a violation of the WEP. Testing of the WEP has a long history, and methods using Newtonian pendulum, torsion balance, free-fall experiment, Lunar Laser Ranging, or astronomical observation have been proposed and implemented to search for the boundary where the WEP is still valid. [4-8] The highest precision of the WEP test on the ground comes from the results of the torsion balance and Lunar Laser Ranging experiments. Both of those experiments had a precision of $10^{-13}$. [5,6]

Atom interferometers provide a new method for testing the WEP using microscopic atoms. By measuring the differential acceleration of the two atomic clouds using dual-species AI, a high-precision WEP test can be realized. The first results of the WEP test based on AI came from *S. Fray et al*. in 2004, they measured the gravitational accelerations of $^{85}$Rb and $^{87}$Rb atoms and obtained a test precision of $10^{-7}$. [9] Subsequently, various AI-based WEP test experiments were carried out. [10-17] The test mass covers rubidium, potassium, strontium, etc. The highest test precision was obtained by *P. Asenbaum et al*. in 2020, they carried out an AI-based WEP experiment using ultracold atoms of the Rb isotope, and obtained a test precision of $10^{-12}$. [18] Testing the WEP using AI does not only improve the test precision but also extend the test category. This is because microscopic atoms can be prepared in different quantum states, such as spin or superposition states. [19-21] The WEP test using these atomic states is a direct quantum test of the theory of gravity.

The microgravity environment has many advantages that the WEP test experiments of both macroscopic and microscopic test masses can benefit from. For macroscopic objects, the French Space Agency has carried out a satellite project called MICROSCOPE. [22] By measuring the differential accelerations of two concentric cylinders of platinum and titanium alloy, a WEP test precision of $10^{-15}$ had been achieved, [22] and it is the highest test precision of the WEP up to now. For microscopic atoms, owing to zero gravity, atom interferometer in space can achieve a long interference time, thus significantly improving the measurement precision. Many AI-based WEP test projects in space have been proposed, including QUANTUS, [23] MAIUS, [24,25] ICE, [26,27] STE-QUEST, [28] etc. The QUANTUS is a drop-tower project which is pre-research for satellite project. It has achieved $^{87}$Rb MOT, [29] $^{87}$Rb BEC [30] and Mach-Zehnder interferometer. [31] The MAIUS is a sounding-rocket project. It has successfully prepared BEC [24] and verified the coherence of BEC in microgravity. [25] ICE is a parabolic flight project, it has obtained the Ramsey fringes and realized a WEP test with precision of $10^{-4}$ using $^{87}$Rb and $^{39}$K atoms. [27] The STE-QUEST project is a satellite project proposed by the European Space Agency (ESA). Its scientific objective is to test the WEP using $^{87}$Rb and $^{41}$K atoms and achieve a test precision at the level of $10^{-17}$. [28] In addition to testing the WEP, AI in space can be also used to detect the gravitational wave and dark energy, projects like Q-WEP, [32] QTEST, [33] BECCAL, [34-36] SAI, [37] SAGE [38] have been proposed.

The China Space Station (CSS) will be in orbit for decades. Many scientific laboratory cabinets (SLCs) are arranged on the CSS to carry out various scientific experiments in space. The microgravity scientific laboratory cabinet (MSLC) is one of the SLCs, and it can achieve a microgravity level of $10^{-7}$ g, which is much better than the CSS itself. This provides an ideal platform for scientific experiments that require a high microgravity. However, limited by the space and power supply of the cabinet, strict restrictions are imposed on the payload design.

Under such circumstances, we designed and achieved a compact dual-species AI payload that can work in MSLC. The scientific objective is to realize a WEP test with high precision. The payload uses rubidium isotopes as the test mass. The overall size of the payload is 460 mm × 330 mm × 260 mm, its weight is approximately 37 kg, and its peak power consumption is approximately 70 W. The payload was launched into the CSS by the TianZhou-5 spacecraft (TZ-5) on November 12, 2022.

The remainder of this paper is organized as follows. In Section 2, we introduce the constraints and guidelines of AI payload design. In Section 3, we introduce the detailed design of the scientific payload, including its composition and functions. In Section 4, we introduce the scientific experimental process in space. In Section 5, we analyze the expected test precision of the WEP test. In Section 6, we present the experimental results of the ground test.

## 2 Key technologies of the payload design

The scientific payload is installed in a magnetic suspension bench (MSB) which is inside the MSLC. The first problem we have to solve is finding a proper scheme that can achieve a relatively high test-precision and fit the design constraints.

### 2.1 Three-dimension velocity selection for achieving ultracold atoms

The most important advantage of microgravity for AI is that it can significantly increase the interference time. However, a long interference time implies a low expansion rate of the atomic cloud, which equivalently implies a low effective atomic ensemble temperature. For rubidium, if one wants the atomic cloud to expand to 1 cm (full width at half maximum) within 2.5 s, then the equivalent temperature is calculated to be 29 nK.

The usual way for obtaining such a low temperature is to use the evaporative cooling method. However, both the magnetic trap and optical trap of the evaporative cooling scheme require a large amount of power consumption, which is far beyond the rated power consumption of the payload design. Therefore, we propose a three-dimensional (3D) velocity selection method to obtain an effective low atomic ensemble temperature. The basic idea is as follows: after the 3D-MOT cooling and polarization gradient cooling (PGC) processes, the atomic cloud expands freely, the velocity of the atomic cloud in the direction of the Raman laser is selected by Raman pulses with proper duration, and the velocities of the atomic cloud perpendicular to the direction of the Raman laser are selected by the detection area. Using this method, we can select the desired 3D atom velocity distribution at the cost of losing a certain number of atoms. This method employs only the traditional 3D cooling technique and does not require additional power consumption.

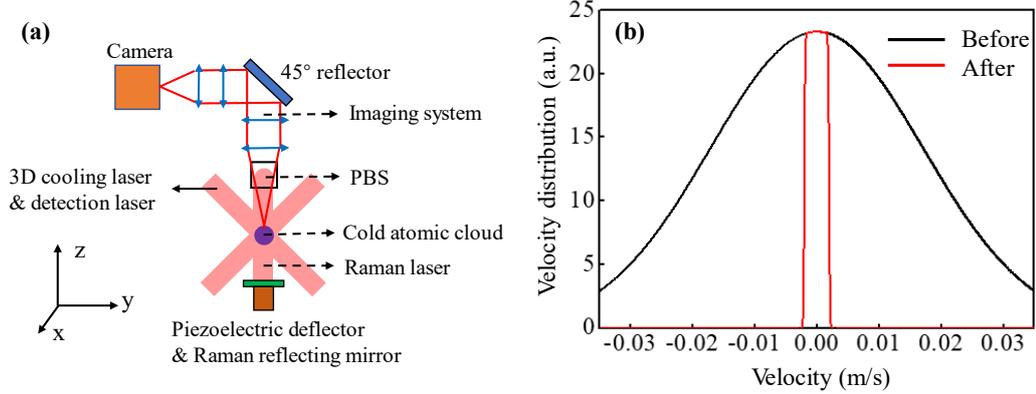

Fig. 1 Scheme of the velocity selection and velocity distribution. (a) Schematic diagram of the 3D velocity selection. (b) 1D velocity distribution of the atomic cloud before and after velocity selection. The number of atoms and temperature of the atomic cloud are $N = 5\times10^8$, $T_e = 4$ μK before the velocity selection, and the parameters for the velocity selection are $r_0 = 0.5$ mm, $l = 0.5$ cm and $t = 2.5$ s.

The experimental scheme of the velocity selection method and the corresponding coordinate frame are shown in Fig. 1. The cold atomic clouds after 3D cooling are localized in the center of the vacuum chamber, the Raman laser is reflected by a polarization beam splitter (PBS) and acts on the cold atomic clouds in the $z$-direction. After the Raman interference process, the fluorescence of the atomic clouds is excited by the 3D cooling laser beams, which propagate along the $x$-axis and in the $y$-$z$ plane.

The fluorescence is transmitted through the PBS and a four-lens imaging system and is imaged by a SCOMS camera. The velocity distribution of the atomic cloud in the z-direction is selected by the Raman laser pulse with a proper duration, and the velocity distribution of the atomic cloud in the x- and y- direction can be selected by the imaging range of the camera.

First, we consider the velocity selection scheme in the one-dimensional (1D) case. Assuming that the temperature after the PGC process is $T_e$, the average atomic cloud velocity and position are zero. The initial widths of the velocity and size distributions of the atomic cloud are $v_0$ and $r_0$, respectively, with $v_0 = \sqrt{k_B T_e / m_{Rb}}$, where $k_B$ is the Boltzmann constant, and $m_{Rb}$ is the mass of Rb atoms. The atomic cloud distribution after an evolving time $t$ is

$$F(r,v,t) = \frac{1}{\sqrt{2\pi} r_0} e^{-(r-vt)^2/2r_0^2} \frac{1}{\sqrt{2\pi} v_0} e^{-v^2/2v_0^2}. \tag{1}$$

The selection range is $(-l, l)$, the velocity distribution is $f(v,t) = \int_{-l}^{l} F(r,v,t)dr$, the selection ratio of the atom number is $\beta_{1D} = \int_{-\infty}^{\infty} f(v,t)dv$, and the effective temperature of the selected atomic cloud is $T_{sel-1D} = \int_{-\infty}^{\infty} m_{Rb} v^2 f(v,t) dv / (\beta_{1D} k_B)$. When the distributions and selection ranges of the atomic cloud are isotropic, the 3D effective temperature and selection ratio are $T_{sel-3D} = 3 T_{sel-1D}$ and $\beta_{3D} = \beta_{1D}^3$.

For our typical experimental parameters, we have $N = 5\times10^8$, $T_e = 4$ μK, $r_0 = 0.5$ mm, $l = 0.5$ cm and $t = 2.5$ s. Then the calculated selection ratio is $\beta_{3D} = 5.2\times10^{-4}$, and selected number of atoms is $N_{sel} = 2.6\times10^5$. The effective atomic temperature is $T_{sel-1D} = 14$ nK. The 3D velocity selection method decreases the atomic cloud temperature by reducing the number of atoms. Therefore, to supply a sufficient number of atoms, the two-dimensional $^+(2D^+)$ [39] scheme was applied in the design of the payload.

### 2.2 Double-diffraction Raman interference scheme associated with the phase shear method

In the following of the article, we denote $i = 1, 2$ for the $^{85}$Rb and $^{87}$Rb atoms and the transition means the D$_2$ line transition for the rubidium atom for simplicity, and F and F′ denote the hyperfine structure of the 5S$_{1/2}$ ground state and 5P$_{3/2}$ exited state, respectively.

In microgravity, the atomic cloud is at rest if its released velocity is zero. When the atomic cloud interacts with a Raman laser or Bragg laser, because of the degenerate of the energy states $|a_i, +\hbar\mathbf{k}_{eff,i}\rangle$ and $|a_i, -\hbar\mathbf{k}_{eff,i}\rangle$, $|a_i\rangle$ and $|b_i\rangle$ represent the $|F=2, m_F=0\rangle$ and $|F=3, m_F=0\rangle$ ground states for $^{85}$Rb ($|F=1, m_F=0\rangle$ and $|F=2, m_F=0\rangle$ for $^{87}$Rb) as shown in Fig. 2, the double-diffraction process occurs spontaneously, where $\mathbf{k}_{eff,i} \equiv \mathbf{k}_{1,i} - \mathbf{k}_{2,i} = \mathbf{k}_{3,i} - \mathbf{k}_{4,i}$ is the effective wave vector, and $\mathbf{k}_{m,i}$ ($m = 1, 2, 3, 4$) is the wave vector. For our design, we choose the double-diffraction Raman scheme[40] for the following reasons. First, the Raman transition is a type of internal state transition, which puts a lower requirement for the atomic temperature than the Bragg transition. Second, the end-state population of the interference can be detected by different internal states, which does not require a space separation fluorescence detection method. Third, the Raman laser was much easier to produce for our phase-modulation based optical system.

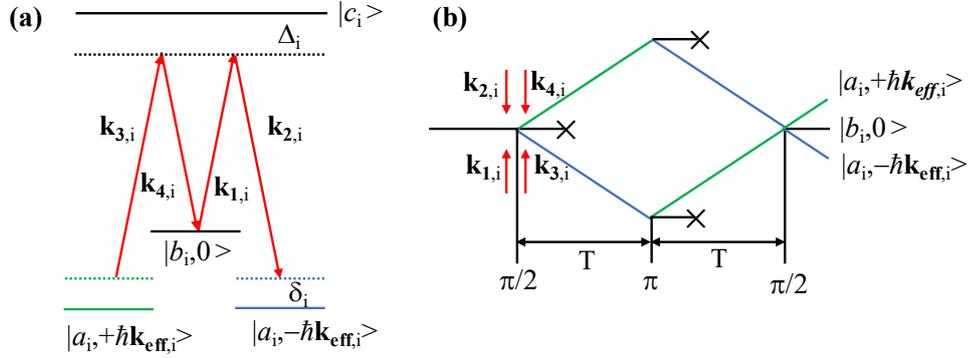

Fig. 2 Schematic diagram of the double-diffraction Raman interference. (a) Energy levels of the double-diffraction Raman transitions. (b) Interference loop of the double-diffraction Raman interference. $|a_i\rangle$ and $|b_i\rangle$ represent the ground states of $|F=2, m_F=0\rangle$ and $|F=3, m_F=0\rangle$ for $^{85}$Rb ($|F=1, m_F=0\rangle$ and $|F=2, m_F=0\rangle$ for $^{87}$Rb), $|c_i\rangle$ is the excited state, $\Delta_i$ and $\delta_i$ are the single-photon and double-photon detuning.

The phase shear method is an effective way to extract the interference fringe. [41] It has advantages of obtaining interference fringes in a single measurement, eliminating noise from atom number fluctuations and detection laser, and greatly suppressing the dephasing effect caused by the rotation and gravity gradient. However, there are still some challenges in applying this method for our experiment, the main reason is the uncertainty of the rotational velocity of the MSB.

When the MSB is in the levitation mode, the rotation of the bench is influenced by the magnetic field and gravity gradient of the Earth. In addition, the noise and zero-point drift of its inertial measurement unit (IMU) gyroscope also result in residual control noise of the rotation. This causes uncertainty in the rotation, both for the magnitude and direction. The phase shear induced by the rotation is

$$\phi_{\text{Rot},i} = 4k_{eff,i}\Omega_x v_{y,i}T^2 - 4k_{eff,i}\Omega_y v_{x,i}T^2, \tag{2}$$

where $\Omega_x$ and $\Omega_y$ are the rotation components of the bench, $v_{x,i}$ and $v_{y,i}$ are the velocity components of the atoms, $T$ is the interference evolution time. Therefore, the direction of the interference fringe of the atomic cloud is always in the *x-y* plane. The best strategy is to detect the fringe in the *z*-direction, which coincides with the scheme of the 3D velocity selection method, as shown in Fig. 1.

In the detection stage, the Raman laser is turned off and a cooling laser is used to excite the fluorescence of the atomic cloud. A PBS is used to separate the Raman laser and fluorescence because they have to be in the same direction. The use of the PBS results in losing of half of the fluorescence intensity.

### 2.3 Rotation compensation scheme with a piezoelectric deflector

The MSB has a weak control of the rotation with a precision of about 0.5 mrad/s. This can result in a fringe numbers variation of 0 to 20 over a 1 s interference time interval and a 1 cm detection range according to Eq. (2). If the rotation rate is close to zero, the phase shear of the atomic cloud will be too small to extract the phase shift, and if the rotation rate is too large, there will be too many fringe numbers in the limited detection range, which will decrease the contrast of the fringe owing to the initial size of the atomic cloud and the resolution of the imaging system.

In order to solve the problem, we use a closed-loop controlled piezoelectric deflector associated with an IMU module to compensate the residual rotation, the Raman laser reflecting mirror is mounted on the deflector as shown in Fig. 1. The tilted angles of the deflector can be controlled during the Raman process according to the rotation measured by the IMU. Considering the tilt angles of the deflector, the phase shift in Eq. (2) is altered

$$\phi_{\text{Rot},i} = 4k_{\text{eff},i}\Omega_x v_{y,i}T^2 + 2k_{\text{eff},i}\theta_{1x}y_i + 2k_{\text{eff},i}\theta_{3x}(y_i + 2v_{y,i}T) \\ - 4k_{\text{eff},i}\Omega_y v_{x,i}T^2 - 2k_{\text{eff},i}\theta_{1y}x_i - 2k_{\text{eff},i}\theta_{3y}(x_i + 2v_{x,i}T) \quad , \tag{3}$$

where $\theta_{n\alpha}$ (n = 1, 3, $\alpha$ = x, y) is the tilt angle component of the deflector for the *n*th Raman pulse and $\alpha_i$ and $v_{\alpha,i}$ are the position and velocity components of the atomic cloud for the first Raman pulse. By properly adjusting $\theta_{n\alpha}$, one can eliminate the decoherence effect caused by the initial position and velocity [18], thus increase the contrast of the interference fringe.

### 2.4 Optimized the Raman laser ratio to eliminate the AC-Stark shift

The Raman lasers cause an AC-Stark shift of the interference transitions of the Rb atoms, which results in phase shifts of the interference fringes. By adjusting the offsets and intensity ratios of the Raman lasers, the AC-Stark shift can be eliminated. However, in our experimental setup, the strategy was slightly more complicated. There are two main reasons. First, the dual-species interferometers are operated synchronously, we have to eliminate the AC-Stark shifts for $^{85}$Rb and $^{87}$Rb simultaneously. Second, the Raman laser sidebands are generated by fiber electro-optical modulators (FEOMs), FEOM produces symmetrical laser sidebands, which makes the calculation of elimination more complex. The differential AC-Stark shifts $\Delta v_{ac,i}$ and the two effective Rabi frequencies $\Omega_{eff,i}$ of the Rb isotope can be expressed as

$$\Delta v_{ac,i} = \sum_{k,j} \frac{\left|\Omega_{k,j,a,i}\right|^2}{4\Delta_{k,j,a,i}} - \frac{\left|\Omega_{k,j,b,i}\right|^2}{4\Delta_{k,j,b,i}}, \tag{4}$$

$$\Omega_{eff,1} = \sum_j \left(\frac{\Omega_{1,j,a,1}\Omega_{2,j,b,1}}{2\Delta_{1,j,a,1}} - \frac{\Omega_{2,j,a,1}\Omega_{3,j,b,1}}{2\Delta_{2,j,a,1}}\right), \tag{5a}$$

$$\Omega_{eff,2} = \sum_j \left(\frac{\Omega_{4,j,a,2}\Omega_{5,j,b,2}}{2\Delta_{4,j,a,2}} - \frac{\Omega_{5,j,a,2}\Omega_{6,j,b,2}}{2\Delta_{5,j,a,2}}\right), \tag{5b}$$

where $\Omega_{k,j,a(b),i}$ and $\Delta_{k,j,a(b),i}$ are the Rabi frequency and single-photon detuning of the Raman transition, $k$ =1, 2…6 represents the label of the Raman lasers, the value 1, 2, 3 (4, 5, 6) represent the +1 order sidebands, carrier and −1 order sidebands of the output lasers of the FEOM for $^{85}$Rb ($^{87}$Rb) manipulation, $j$ denotes the label of available excited state for the Rb Raman transition, and $a$ and $b$ represent the $^{85}$Rb |$F$=2, $m_F$=0⟩ and |$F$=3, $m_F$=0⟩ ($^{87}$Rb |$F$=1, $m_F$=0⟩ and |$F$=2, $m_F$=0⟩) ground states.

To determine the optimized intensity ratios and frequency offsets of the Raman lasers, we set the following constraints. First, both the differential AC-Stark shifts of $^{85}$Rb and $^{87}$Rb are zero, that is $\Delta v_{ac,1} = \Delta v_{ac,2} = 0$, and the effective Rabi frequencies of $^{85}$Rb and $^{87}$Rb are equal, that is, $\Omega_{eff,1} = \Omega_{eff,2}$. We only consider the carrier and ±1 order sidebands for both $^{85}$Rb and $^{87}$Rb Raman lasers. Then, there are five variables that we can optimize: the two frequency offsets of the two carriers, the two first-order sideband-to-carrier intensity ratios of the $^{85}$Rb and $^{87}$Rb Raman lasers, and the intensity ratio of the $^{85}$Rb Raman laser carrier to the $^{87}$Rb Raman laser carrier. The number of variables is greater than the number of equations. Therefore, the optimized variables are not unique. We provide some other constraints for the variables: the single-photon detuning is as large as possible to avoid spontaneous emission, the first-order sideband-to-carrier intensity ratio is not too large to avoid high-order sidebands. We found a following set of optimized parameters. The frequency offset of the $^{85}$Rb Raman laser carrier is 986 MHz blue detuned to the $^{85}$Rb |$F$=3⟩→|$F'$=4⟩ transition, and the $^{87}$Rb Raman laser carrier is 404 MHz blue detuned to the $^{87}$Rb |$F$=2⟩→|$F'$=3⟩ transition, and the intensity ratio of the $^{85}$Rb Raman laser +1 order sideband and its carrier, the $^{87}$Rb Raman laser

+1 order sideband and its carrier is 1:1.34:1.69:0.63.

## 3 Design of the space cold atom interferometer payload

The space cold atom interferometer payload contains three systems: physical, optical, and electronic systems. The three systems were assembled into a 460 mm × 330 mm × 260 mm cabinet.

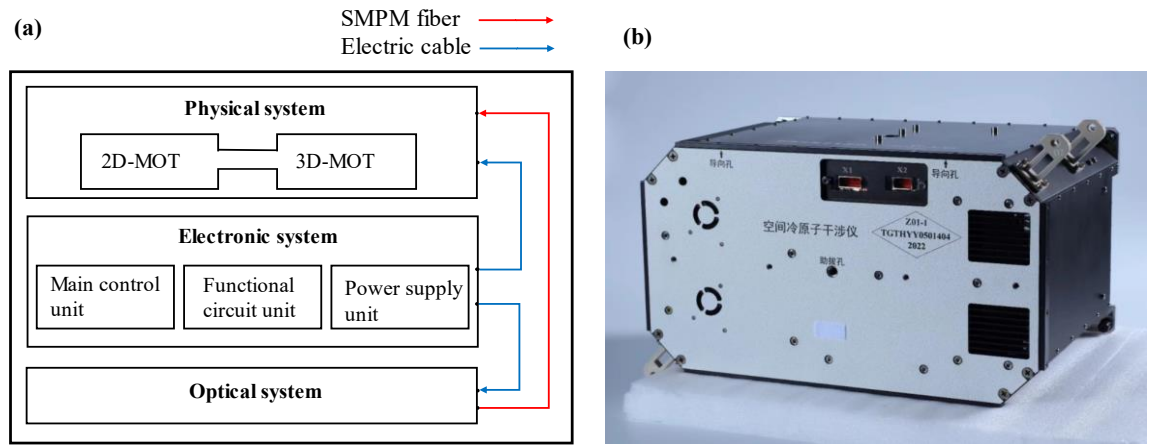

Fig. 3 The space cold atom interferometer payload. (a) Schematic diagram of the scientific payload. (b) Picture of the scientific payload.

### 3.1 Design of the physical system

The physical system is the place where laser cooling and atom interference occur. It contains a vacuum chamber, its accessories, the supporting structures, the optical components, and a magnetic shield, as shown in Fig. 4.

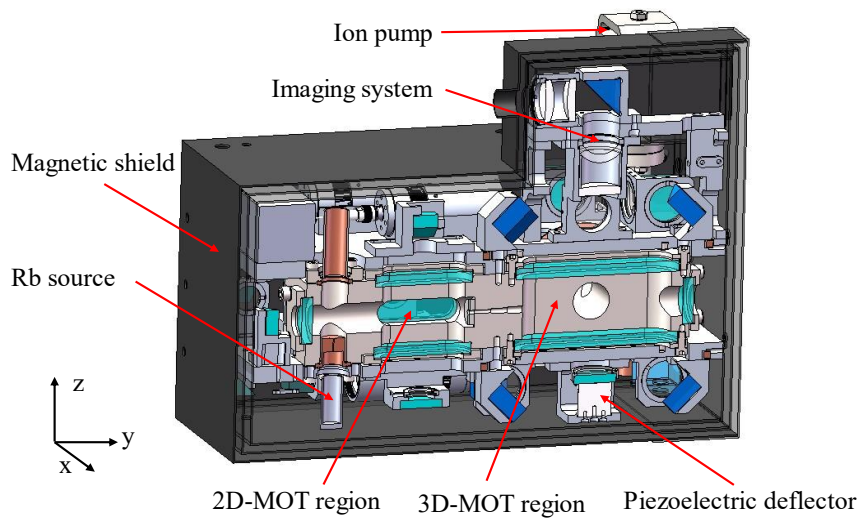

Fig. 4 Cross section view of the physical system

The vacuum chamber is made of titanium alloy and is maintained by a 2 $l$ ion pump and a getter pump with a level of $10^{-8}$ Pa. The chamber has two regions. The 2D-MOT region is used to generate a high-flux cold atom beam. It has a rubidium source with a heating unit to control the amount of released rubidium. 2D-MOT has a 2D$^+$-MOT configuration. A circularly polarized laser beam in the *y*-direction passes through the chamber and is reflected by a reflecting mirror to form the push beam of the 2D$^+$-MOT. The reflecting mirror is mounted inside the chamber and has a hole with a diameter of 2 mm. A $\lambda/4$ wave plate with a 2 mm hole is attached in front of the mirror. The pushing beam, together with two pairs of laser beams perpendicular to it, form the cooling lasers of the 2D$^+$-MOT, and a pair of anti-Helmholtz coils is used to provide the magnetic field gradient for the 2D$^+$-MOT. The cooled cold atom beam reaches the 3D-MOT region through the hole of the reflecting mirror and a channel that connect the two regions.

In the 3D-MOT region, one laser beam is reflected by a series of wave plates and reflecting mirrors to form three pairs of cooling laser beams. These beams are in the *x* direction and in the *y*-*z* plane. A pair of anti-Helmholtz coils is used to provide a magnetic field gradient for the 3D-MOT. The cooled dual-species Rb atoms are localized at the center of the 3D-MOT chamber. Another laser beam is reflected by the PBS and propagates in the *z*-direction, which acts as the Raman laser. The laser beam passes through the cold atoms and is reflected by a reflecting mirror mounted on a piezoelectric deflector. A $\lambda/4$ wave plate is attached to the reflecting mirror. The fluorescence of cold atoms in the *z*-direction passes through the PBS and a four-lens imaging system and is imaged by a SCOMS camera. Three photodetectors were installed to detect the power of the 2D, 3D, and Raman lasers.

Three layers of permalloy shells are mounted outside the chamber to shield the external magnetic field. A flux-gate magnetometer is installed inside the magnetic shielding shells to detect residual magnetic fields. Several electronic components are also mounted outside the shells, including the driver for the ion pump, the driver for the piezoelectric deflector, the photon current amplifying circuit, and an IMU module.

The physical system has a size of 300 mm × 200 mm × 232 mm and a weight of 16 kg. The laser for the physical system is transmitted from the optical system through three single-mode polarization-maintaining fibers.

### 3.2　Design of the optical system

To minimize the volume and increase the thermostability of the optical system, we use a fused quartz glass plate as the baseboard of the optical system. Miniaturized optical components are installed on both sides of the board.

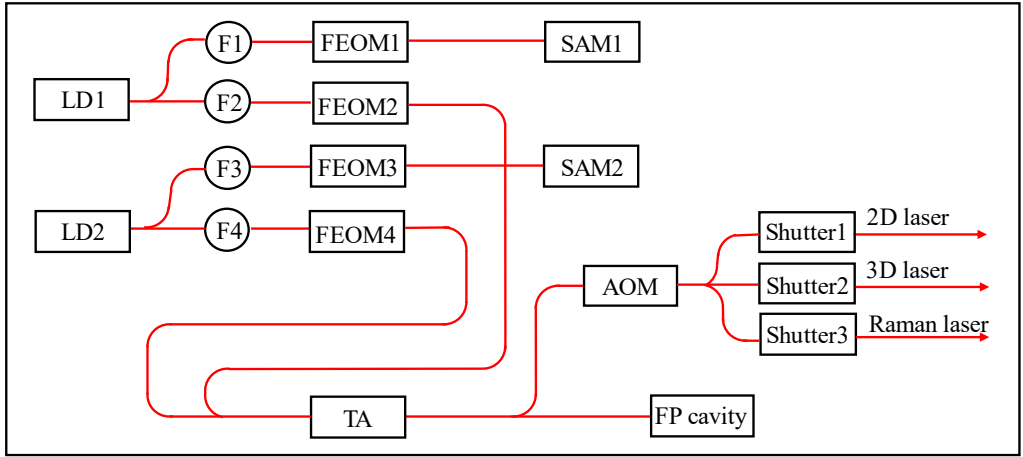

Fig. 5 Schematic diagram of the optical system

The principle of the optical system is shown in Fig. 5. Two DFB laser diodes (LDs) are used to generate lasers for $^{85}$Rb and $^{87}$Rb manipulation, labeled as LD1 and LD2, respectively. The output laser of LD1 is split in two ways and then coupled to two single-mode polarization-maintaining (SMPM) fibers, labeled as F1 and F2. F1 is connected to FEOM1, and then sent to a saturation absorption module (SAM), labeled as SAM1. SAM1 uses the Rb isotope 87 vapor cell. And the −1 order sideband of FEOM1 is lock to $^{87}$Rb $|F=2\rangle \rightarrow |F'=co(2, 3)\rangle$ transition. By varying the frequency of microwave that drives FEOM1, we can change the carrier frequency of LD1 from resonance to $^{85}$Rb $|F=3\rangle \rightarrow |F'=4\rangle$ transition to blue-detuning to this transition for several hundred MHz. F2 is connected to FEOM2, which is driven by microwaves ranging from 2.7 to 3.6 GHz, and the carrier and sidebands of the FEOM2 output laser act as $^{85}$Rb cooling, Raman, and detection lasers. The output laser of LD2 has an optical path similar to that of LD1. The difference is that SAM2 employs a $^{85}$Rb vapor cell, the +1 order sideband of FEOM3 is locked to the $^{85}$Rb $|F=3\rangle \rightarrow |F'=co(3, 4)\rangle$ transition, and the driven frequency of FEOM4 is ranging from 5.0 to 6.9 GHz.

The outputs of FEOM2 and FEOM4 are combined and injected into a tapered amplifier (TA) to amplify the output power. A few fractions of the TA output laser power are split and coupled to a Fabry–Pérot (FP) cavity. The cavity is used to monitor the powers of the carriers and sidebands of the two amplified lasers. Most of the TA output laser power is diffracted by an acousto-optic modulator (AOM) and split into three paths. In each path, a mechanical shutter controls the opening or closing of the output laser with a high extinction ratio. The lasers of the three paths are then coupled to three SMPM fibers and sent to the physical system. They act as 2D, 3D, and Raman lasers. The maximum total output power is about 100 mW. Several photodetectors are installed in the optical system to monitor the laser powers of the nodes of the optical path. The optical system has a size of 270 mm × 110 mm × 250 mm and a weight of 5.2 kg.

### 3.3 Design of the electronic system

The electronic system is used to drive and control the payload. It is divided into three units: the power supply, main control, and functional circuit units. The payload operates in remote control mode. All operations can be only realized by commands sent from the ground. The MSLC receives these commands and transfers them to the scientific payload.

The power supply unit is used to convert the 28 V input voltage to ±5, 12, ±15, 24, and 28 V output voltages, which are used to provide the power supplies for the entire electronic system. The power supply unit also has a power switch function; it can receive commands from the main control unit to control the switches of the output power supply. The main control unit has several functions, including receiving and processing the transferred commands, controlling the switch states of the power supplies, setting the parameters and switch states of the functional circuits, controlling scientific processes such as laser cooling and atom interference by generating AO and DO time sequences, acquiring and processing the generated data, and transferring them to the ground. The functional circuit unit is a collection of a series of individual circuits, including a constant-current circuit, temperature control circuit, and microwave generation circuit. A constant-current circuit is used to drive the LDs, TA, shutters, and coils in the physical system. The temperature control circuit is used to control the temperature of the LDs, TA, Rb vapor cells, and the rubidium source of the vacuum chamber. The microwave generation circuit is used to generate microwaves to drive the four FEOMs and the AOM. The electronic system had a stacked structure with a size of 247 mm × 140 mm × 250 mm and a weight of 6.5 kg.

### 4 Experiment process in space

The major scientific goal of our project is to test WEP with high precision. To achieve this goal, the experimental process was divided into the laser cooling, atom interference, and fluorescence detection stages.

The laser cooling stage can be further divided into 2D-MOT, 3D-MOT, and PGC stages. In the 2D-MOT stage, the shutters for the 2D and 3D lasers are open and the shutter for the Raman laser is closed, the carrier frequency of LD1 (LD2) is set to be −16 MHz detuning to the $^{85}$Rb $|F=3\rangle \rightarrow |F'=4\rangle$ transition ($^{87}$Rb $|F=2\rangle \rightarrow |F'=3\rangle$ transition), and the +1 sidebands of LD1 (LD2) are in resonance with the $^{85}$Rb $|F=2\rangle \rightarrow |F'=3\rangle$ transition ($^{87}$Rb $|F=1\rangle \rightarrow |F'=2\rangle$ transition). The anti-Helmholtz coils for the 2D$^+$ and 3D-MOT were turned on. In this state, a slow cold atom beam is generated in the 2D-MOT region and transferred to the 3D-MOT region, in which 3D-MOT accumulate atoms for several seconds. In the 3D-MOT stage, the shutter for the 2D laser was closed, and the coils for the 2D$^+$-MOT were turned off. The carrier frequency of LD1 (LD2) was set to be −60 MHz detuning to the $^{85}$Rb $|F=3\rangle\rightarrow|F'=4\rangle$ transition ($^{87}$Rb $|F=2\rangle\rightarrow|F'=3\rangle$ transition), whereas the +1 sidebands of LD1 (LD2) were still in resonance with the $^{85}$Rb $|F=2\rangle\rightarrow|F'=3\rangle$ transition ($^{87}$Rb $|F=1\rangle\rightarrow|F'=2\rangle$ transition). This stage lasts for approximately 100 ms, because the influence of the cold atom beam and the magnetic field of the 2D-MOT are absent, the position of the 3D-MOT can be localized to the center of the 3D-MOT chamber. In the PGC stage, the coils for the 3D-MOT were turned off, the intensity of the cooling laser was turned to zero in approximately 4 ms, and the temperature was decreased to several μK.

In the atom interference stage, the shutter for the 3D laser was closed, and the shutter for the Raman laser was open. The carrier frequency of LD1 (LD2) was set to be 986 MHz (404 MHz) detuning to the $^{85}$Rb $|F=3\rangle\rightarrow|F'=4\rangle$ transition ($^{87}$Rb $|F=2\rangle\rightarrow|F'=3\rangle$ transition), and the frequency of the microwave for FEOM2 (FEOM4) was tuned to 3.035 GHz (6.835 GHz). The Raman laser was turned on for tens of μs for the first Raman pulse, and the $^{85}$Rb ($^{87}$Rb) atoms are transferred from the $|F=3, m_F=0\rangle$ ($|F=2, m_F=0\rangle$) state to the $|F=2, m_F=0\rangle$ ($|F=1, m_F=0\rangle$) state. Then, the carrier frequency of LD1 (LD2) was set to be in resonance with the $^{85}$Rb $|F=3\rangle\rightarrow|F'=4\rangle$ transition ($^{87}$Rb $|F=2\rangle\rightarrow|F'=3\rangle$ transition), and the sidebands for LD1(LD2) were turned off by shutting down the corresponding microwave. The blow-away laser was turned on for tens of μs, and this process removed the remaining $^{85}$Rb ($^{87}$Rb) atoms in the $|F=3\rangle$ ($|F=2\rangle$) state. Subsequently, the second Raman pulse, the second blow-away pulse, and the third Raman pulse were turned on in sequence. After the interference sequence, the $^{85}$Rb ($^{87}$Rb) atoms were in superposition states of $|F=2, m_F=0\rangle$ and $|F=3, m_F=0\rangle$ ($|F=1, m_F=0\rangle$ and $|F=2, m_F=0\rangle$). In the interference stage, the angle of the deflector at the moments of the three Raman pulses was turned according to the rotation rate measured by the IMU to compensate for the residual rotation. The maximum interval between the Raman pulses was approximately 1 s, which is limited by the size of the vacuum chamber.

In the fluorescence detection stage, the shutter for the 3D laser was open, the shutter for the Raman laser was closed, the carrier frequency of LD1 (LD2) was set to be in resonance with the $^{85}$Rb $|F=3\rangle\rightarrow|F'=4\rangle$ transition ($^{87}$Rb $|F=2\rangle\rightarrow|F'=3\rangle$ transition), and the sidebands for LD1 (LD2) were turned off by shutting down the corresponding microwave. We turned on the detection pulses for $^{85}$Rb and $^{87}$Rb in sequence, and the fluorescence of the $^{85}$Rb ($^{87}$Rb) atoms in the $|F=3, m_F=0\rangle$ ($|F=2, m_F=0\rangle$) were imaged by the camera. The minimum time interval between the two detection pulses was approximately 20 ms, which was limited by the frames per second (FPS) of the camera. From the image of the atomic cloud, we can obtain the phase shear interference fringe for the atomic cloud, where the phase of the fringe represents the acceleration felt by the atomic cloud. By comparing the accelerations of the two atomic clouds, we can test the WEP.

## 5 Analysis of the test precision

The breaking of the WEP can be expressed by the Eötvös coefficient $\eta$

$$\eta = \frac{g_1 - g_2}{(g_1 + g_2)/2} = \frac{(g_1 - a_s) - (g_2 - a_s)}{(g_1 + g_2)/2} = \frac{a_1 - a_2}{(g_1 + g_2)/2} = \frac{\Delta a}{g}, \tag{6}$$

where $g_i$ is the gravitational acceleration for the Rb isotope. For experiment in space, the atomic cloud feels a centrifugal force $a_s$ in Newton's laws. The measured atom acceleration is $a_i$ in the frame of the aircraft; thus, $\eta$ can be measured by the differential acceleration $\Delta a$ and the gravitational acceleration g of the earth, as illustrated by Eq. (6). The orbital altitude of the CSS is approximately 400 km, and the corresponding gravitational acceleration is g = 8.7 m/s$^2$. The phase of the atom interference fringe induced by the acceleration is $\phi_i = 2k_{\text{eff},i} a_i T^2$. Then $\eta$ can be expressed as

$$\eta = \frac{\Delta\phi}{2k_{\text{eff},1} g T^2} - \frac{\Delta k}{2k_{\text{eff},1}} \frac{a_2}{g}, \tag{7}$$

where $\Delta\phi = \phi_1 - \phi_2$ is the differential phase and $\Delta k = 2(k_{\text{eff},1} - k_{\text{eff},2})$ is the differential wave vector.

For the first term in Eq. (7), the parameters $k_{\text{eff},1}$, $T$ can be set by the experiment, and g can be calculated by the orbit of CSS. The normalized precision of the setting value ($k_{\text{eff},i}$ to 10$^{-8}$, $T$ to 10$^{-6}$, and g to 10$^{-3}$) are much higher than the aimed normalized precision (approximately 1 if WEP does not break) of $\eta$ to be measured. Thus, the values of $k_{\text{eff},i}$, $T$ and g can be safely used in this term without loss of measurement precision. The second term has a suppression ratio term $\Delta k/2k_{\text{eff},1}$, which was 5×10$^{-7}$ in our experiment, and a relative acceleration term $a_2/g$. If one wants the value of this term to be less than 10$^{-11}$, it results in an upper limit for the residual acceleration $a_2$ of 2×10$^{-5}$g, and this can be guaranteed by the MSB. Therefore, the measured precision of $\eta$ is primarily limited by the differential phase $\Delta\phi$.

Many factors influence the value of $\Delta\phi$, and we list the most important terms as below, and we set $T$=1 for the following calculation.

### 5.1 Quantum projection noise

Quantum projection noise originates from the state collapse process of a superposition state. For a dual-species atom interferometer, the differential phase noise caused by the quantum projection effect can be expressed as

$$\sigma_{\Delta\phi} = \sqrt{\sum_{i=1,2} \frac{1}{N_i C_i^2}}, \tag{8}$$

where $N_i$ is the atomic number participating in the interference process, $C_i$ is the contrast of the interference fringe. Considering the magnetic sublevels and consumption of the atom number caused by the velocity selection, the atoms participate in the interference are $N_1 = 3.7 \times 10^4$ and $N_2 = 5.2 \times 10^4$. Assuming that $C_i = 0.2$, the differential phase noise caused by the quantum projection effect in a single shot is 34 mrad.

### 5.2 Detection noise

The image of the interference fringe was captured by the SCMOS camera, and there was detection noise in the imaging process. This noise has three origins: the shot noise, which is a quantum noise, and is relative to the photoelectron number converted by the camera. The readout noise is caused by the noise floor of the camera's readout circuit. The dark noise is the thermal photoelectron noise of the SCMOS. The effective differential phase noise caused by the detection noise can be expressed as

$$\sigma_{\Delta\phi} = \sqrt{\sum_{i=1,2} \left( \frac{1}{C_i^2 N_{counts,i}} + \frac{m\delta_{read}^2}{C_i^2 N_{counts,i}^2} + \frac{mt_{exp}\delta_{dark}^2}{C_i^2 N_{counts,i}^2} \right)}. \tag{9}$$

where $N_{counts,i}$ is the total number of photoelectrons recorded by the camera, $m$ is the number of pixels, $\delta_{read}$ and $\delta_{dark}$ are the readout noise and dark noise of the camera, which are 2.2 e/pixel and 2.3 e/pixel/s$^{1/2}$ for the used camera, and $t_{exp}$ is the exposure time. The total number of photoelectrons was calculated using the following equation,

$$N_{counts,i} = N_i \beta \gamma t_{det} \frac{\Gamma_i I_{0,i} / I_{s,i}}{2(1 + 4(\Delta_i / \Gamma_i)^2 + I_{0,i} / I_{s,i})}, \tag{10}$$

where $\beta$ is the quantum efficiency of the camera, $\gamma$ is the detection ratio of atomic fluorescence, $t_{det}$ is the duration of the detection laser, $\Gamma_i$ is the decay rate for the Rb D$_2$ transition, $I_{0,i}$ is the intensity of the detection laser, $I_{s,i}$ is the saturation intensity of the Rb D$_2$ transition. For the typical parameters that we used, the differential phase noise caused by the detection noise is about 22 mrad in one shot.

### 5.3 Effect of the rotation and gravity gradient

The rotation of the payload and gravity gradient of the Earth and CSS induces a phase shift to the interference fringe. The rotation-induced phase shift is given by Eq. (3), and the gravity gradient-induced phase is

$$\phi_{gg,i} = 2k_{eff,i} T_{zz} z_i T^2 + 2k_{eff,i} T_{zz} v_{z,i} T^3, \tag{11}$$

where $T_{zz}$ is the (z, z) component of the gravity gradient tensor, and $z_i$ and $v_{z,i}$ are the position and velocity components of the atoms for the first Raman pulse. According to Eq. (3) and (11), offsets for the initial position and velocity of the two atomic clouds will induce a differential phase shift. For our payload design, the rotation measurement uncertainty of the IMU is about 1 μrad/s and the angle control uncertainty of the deflector is about 5 μrad, the gravity gradient at the altitude of the CSS has value of about $T_{zz}$ = 2500E. The difference of the initial position and velocity of the two atomic clouds were approximately 200 μm and 100 μm/s according to the test results on ground. Then, the uncertainty differential phases caused by the rotation and gravity gradient were calculated to be 102 and 18 mrad, respectively.

### 5.4 Effect of the magnetic field

In the process of atomic interference, a uniform magnetic field is generated in the z-direction to act as the quantization magnetic field. For the double-diffraction Raman interference transition, the differential phase shift caused by the second-order Zeeman effect of the states is

$$\Delta\phi_{mag} = 4\pi\hbar(k_{eff,1}\alpha_1/m_1 - k_{eff,2}\alpha_2/m_2)B_0\gamma_m T^2, \qquad (12)$$

where $\hbar$ is Planck's constant, $\alpha_i$ is the clock transition Zeeman shift coefficient, $m_i$ is the mass of the Rb atom, $B_0$ is the intensity of the magnetic field intensity, and $\gamma_m$ is the magnetic field gradient. To reduce this phase shift, three layers of magnetic shielding shells were used to attenuate the external stray magnetic field, and the parameters of the interference coils were carefully designed to reduce the magnetic gradient. For our experiment, the magnetic field can be adjusted to 0.1 G and the measured magnetic gradient is less than 3 mG/m, then the differential phase caused by the second Zeeman effect was 33 mrad.

### 5.5 Effect of the AC-Stark shift

We derived an optimized Raman laser ratio to eliminate the differential AC-Stark shift in Section 2.4 However, the AC-Stark shift for a single state still exists. The interference loop of the double-diffraction Raman interference employs only one single internal state, and a constant AC-Stark shift of this state does not cause a fringe phase shift. The phase shift can be induced only by the AC-Stark shift difference $\delta v_{ac,i}$ of the two branches of the interference loop at the π pulse. The induced phase shift can be calculated as follows,

$$\phi_{ac,i} = 4\delta v_{ac,i}\tau, \qquad (13)$$

where $\tau$ is the duration of the π pulse. $\delta v_{ac,i}$ can be induced by the intensity gradient of the Raman laser in the $z$-direction.

For the parameters of our experiment, if we set the duration of the $\pi$ pulse is $\tau = $ 160 μs, then the AC-Stark shift is calculated to be 387.4 Hz (2263.1 Hz) for the $^{85}$Rb atom $|F=2, m_F=0\rangle$ state ($^{87}$Rb atom $|F=1, m_F=0\rangle$ state). The diameter of the Raman laser is designed to be 1.7 cm, for a separation of the atomic cloud wave packet of 2 cm, the relative difference intensity is that the atom felt at the two branches is calculated to be about $8\times10^{-5}$. Then the differential phase shift induced by the AC-Stark shift is about 1 μrad.

### 5.6   Effect of the wave front aberration

The aberration of the reflecting mirror of the Raman laser will causes a phase distortion of the Raman laser in the $x$-$y$ plane. This phase distortion will be coupled with the atomic cloud sizes at the three Raman pulses and induces a phase shift for the interference fringe. [42] For experiment of our case, the phase shifts are mostly in common. This is because that we use the velocity selection method, so the selected atomic clouds have a same expansion rate, and the difference of effective Raman vectors of the Rb atoms is small, this will induce similar phase shifts for the two AIs, as illustrated in the paper. [43] For a mirror with a diameter of 22 mm, and has a peak aberration of $\lambda/5$, the differential phase is calculated to be about 2.3 μrad.

The estimated uncertainties of the WEP test experiment for 100 shots are listed in Table 1. The estimated uncertainties contain both the random uncertainties and the systematic uncertainties. The mainly error source comes from the rotation, the gravity gradient and the magnetic field. The expected test precision is in the order of $10^{-10}$.

Table 1. Estimated test uncertainties of the WEP test experiment

| Error source | Estimated uncertainty |
|---|---|
| Quantum projection noise | $1.2\times10^{-11}$ |
| Detection noise | $8.0\times10^{-12}$ |
| Rotation | $3.6\times10^{-10}$ |
| Gravity gradient | $6.4\times10^{-11}$ |
| Magnetic field | $1.2\times10^{-10}$ |
| AC-Stark shift | $3.4\times10^{-15}$ |
| Wavefront aberration | $1.0\times10^{-14}$ |
| Total | $4.0\times10^{-10}$ |

## 6   Ground test for the scientific payload

We carried out tests to verify the functions and performances of the scientific payload. The tests covered the power supply, communication, functional circuits, optical system, and physical system. We also conducted an environmental test to verify that it can adapt to the environment for rocket launching and in-orbit operation. The payload passed the sinusoidal vibration (5.6 g in peak) and random vibration tests (4.28 g RMS) with the protection of a shockproof package and passed the thermal cycling test from 15 to 32 °C. We also carried out atom manipulation experiments, including atom cooling and atom interference experiments.

For the experiment in space, the cold atomic clouds are at rest in the center of the 3D-MOT chamber after release. However, for the ground-based experiments, the released cold atomic cloud will free fall under the gravity of the Earth. If we set the *z*-axis along the direction of gravity, then the limited detection range allowed only tens of milliseconds of free fall time. This caused problem for both the time-of-flight (TOF) image and the atom interference produced. To extend the allowed detection time, we rotated our payload and set the *x*-axis along the direction of gravity. Under these conditions, the free-fall atom can be detected by the laser in the *x*-direction and is in the object plane of the imaging system. The maximum detectable time was limited by the image range of the camera and was approximately 45 ms in our experiment.

### 6.1 Laser cooling experiment

The laser cooling experiment follows the produce introduced in Section 4. The image of the cold $^{85}$Rb atomic cloud in the 3D-MOT and TOF images of the atomic clouds after the PGC process are shown in Fig. 6. The atom number was calculated to be more than $2\times10^8$ and the calculated atomic temperature was about 4 μK in one dimension.

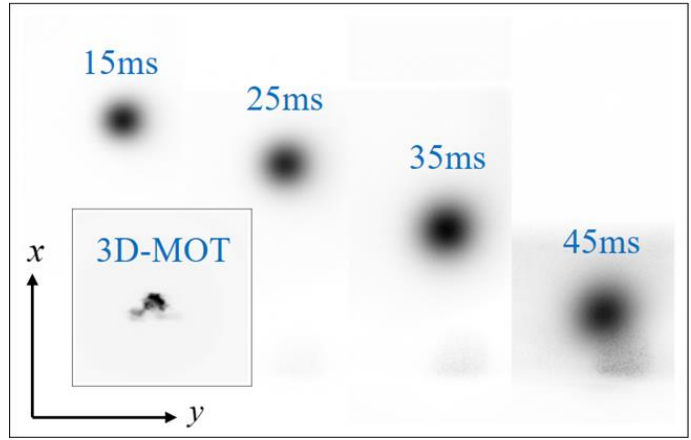

Fig. 6 TOF fluorescence images of the $^{85}$Rb atomic cloud with different falling time. Insert image is 3D-MOT fluorescence image.

### 6.2 Double-diffraction Raman interference experiment

After the PGC process, the shutter for the 3D laser was closed, and the shutter for the Raman laser was open. The carrier frequency of LD1 was set to be resonant with the $^{85}$Rb $|F=3\rangle \to |F'=3\rangle$ transition, the sidebands for LD1 were turned off, the laser was turned on for tens of microseconds, and the $^{85}$Rb atoms were optically pumped from the $|F=3\rangle$ state to the $|F=2\rangle$ state. Then, the carrier frequency of LD1 was set to be 800 MHz blue detuning to the $^{85}$Rb $|F=3\rangle \rightarrow |F'=4\rangle$ transition, and the frequency of the microwave for FEOM2 was tuned to 3.035 GHz. The Raman laser was turned on for 80 μs, 160 μs and 80 μs for the first, second and third Raman pulses, and the interference time interval was set to 5 ms. After the interference sequence, the atoms in the $|F=3\rangle$ state were imaged. To realize the phase shear of an atomic cloud, the angle of the deflector was linearly scanned during the interference to induce a phase shift.

We used principal component analysis (PCA) process to obtain the interference fringe from the atomic interference images. We used 38 images for the process and took the 4th order of the processed images. The curve in Fig. 7 is the average of the processed images in the *x*-direction. To obtain the interference fringe, we tuned the Raman laser to be in resonance, and rotated the deflector at a proper rotation rate. The obtained curve is shown as the blue solid line. We can see a clear fringe. We also carried out comparative experiments. First, we kept the deflector with the same rotation rate, and tuned the Raman laser to be detuned by +100 kHz. The obtained curve is shown as the blue dashed line, the fringe disappeared. Then we stop the deflector rotation, and tuned the Raman laser to be in resonance again. The obtained curve is shown as red solid line. Again, the fringe disappeared.

The contrast of the obtained interference fringe is very low. This is mainly caused by the following reasons. First, during the interference process, the laser intensities of the three Raman pulses were different because the atoms fell in the *x*-direction, and the Raman laser was propagated in the *z*-direction. Second, after the interference process, we detected atoms in the state $|F=3\rangle$. However, during the interference process, most atoms were populated in the $|F=2\rangle$ magnetic sublevels. The single-photon transition effect of the Raman laser pumped a proportion of these atoms to the $|F=3\rangle$ state, and this contributed to the background of the interference fringe. Third, the interference time was too short to set a stable angular sequence for the deflector. We could only scan the angle at a certain rate, therefore, we could hardly optimize the three angles of the deflector for the three Raman pulses, as described in Section 2.3, which induced a decoherence effect caused by the initial size of the atomic cloud. However, all of these factors can be relieved for the experiment in space.

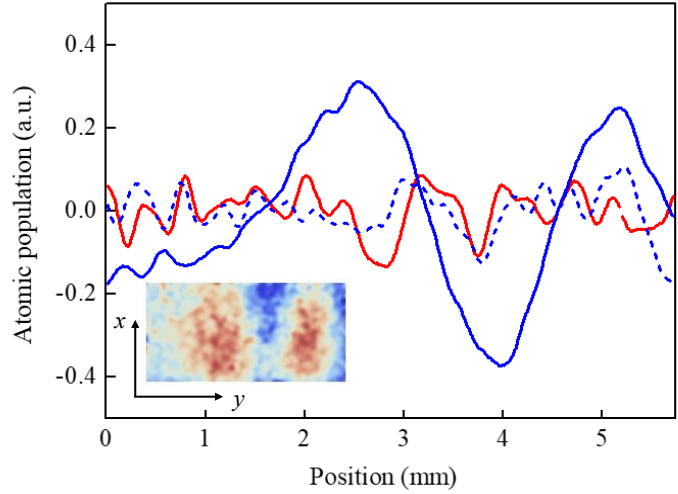

Fig. 7 Double-diffraction Raman interference fringes of the $^{85}$Rb atoms. The PCA process is applied and 38 images are used for each curve. The curves are the averages of the processed images of the 4th order in the *x*-direction. The blue solid line and the insert image represents the case when the Raman laser is in resonance and the deflector rotates at a proper rate. The blue dashed line represents the case when the Raman laser is out of resonance and the deflector rotates at a same rate. The red solid line represents the case when the Raman laser is in resonance and the deflector is not rotated.

## 7  Conclusion

The MSLC in the CSS provides a good environment for experiments that require a high-level of microgravity. In this paper, we designed and realized a dual-species atom interferometer that will work in the MSLC in space, aiming at testing the WEP with high precision. We introduced the scheme of the design of our atom interferometer, which is mainly constructed by the method of 3D velocity selection, the method of double-diffraction Raman interference scheme associated with the phase shear fluorescence detection, and the method of rotation compensation by a piezoelectric deflector. We optimized Raman laser ratio for eliminating the AC-Stark shift for our experiment. Then we designed experimental process for space experiment, the compact scientific payload and its sub-systems. We carefully analyze the test precision of WEP of our payload by the parameters of the pre-researches, the expected test precision is in the order of $10^{-10}$. We also present the laser cooling and interference experiment results on the ground.

The MSLC also restricts the size and power consumption of the scientific payload that it supports. The payload we design is extremely compact and have a size of only 460 mm × 330 mm × 260 mm. This limits the expected test precision for several aspects. The interference time is limited by the size of the vacuum chamber. The interference produce can only begin from atoms that created by the dual-species 3D-MOT, rather than ultra cold atoms that trapped by the magnetic trap or the optical dipole trap. The rotation and compensation angle of payload can not to be controlled to an extremely high level due to the current microgravity environment and the performance to the compact piezoelectric deflector. The gravity gradient compensation strategy cannot be utilized due to the limited frequency detuning range of the optical system. It should be noted that for atom interferometer works on a platform with better microgravity performance and relaxed design restrictions, much higher measurement accuracy can be obtained.

We hope that the design and implementation of the payload can provide certain benefits for future space-based atom interferometer designs.

**Data availability**

The data that support the findings of this study are available from the corresponding author upon request.

**Acknowledgement**

We would like to thank the support from Technology and Engineering Center for Space Utilization, CAS, especially Hongen Zhong, Zongfeng Li, and many others for their constructive discussions and technical supports. We would like to thank professor Yuri Ovchinnikov for his kind guidance for the design of the 2D-MOT. This work was supported by the Second Batch of Scientific Experiment Project of the Space Engineering Application System of the China Space Station. We acknowledge support from the Technological Innovation 2030 "Quantum Communication and Quantum Computer" Major Project (2021ZD0300603, 2021ZD0300604), the Hubei Provincial Science and Technology Major Project (ZDZX2022000001) and National Natural Science Foundation of China (91536221, 12204493).

**AUTHOR CONTRIBUTIONS**

X.C. led the planning, design, development, and testing of the project. J.Q.Z., X.W.Z., Y.W and G.G.G were responsible for the development of physical system. J.F., D.F.Z, and X.L. were responsible for the development of optical system. Q.F.C., H.Y.S, W.H.P, L. Q., Z.Y.X., Z.D.C. and W.Z.W. were responsible for the development of electronic system. K.L., J.T.L., Z.Y.C., G.Z., Y.Z. were responsible for software development. Y.B.W., M.H, J.T.L, L.Z, C.H., R.D.X. and J.J.J. were responsible for the daily experimental operation, payload testing and data acquisition and analysis. M.S.Z. and J.W. proposed the experiment and coordinated with the principal members as the project scientists. M.H. and X.C. prepared the manuscript. All authors have read and approved the final manuscript.

**ADDITIONAL INFORMATION**

Competing interests: The authors declare no competing interests.